\documentclass{article}
\usepackage{epsf}
\usepackage{a4}
\usepackage{geometry}
\usepackage{lscape}
\begin{document}                
\newcommand{\manual}{rm}        
\newcommand\bs{\char '134 }     

\newcommand{\simlt}{\stackrel{<}{{}_\sim}}
\newcommand{\simgt}{\stackrel{>}{{}_\sim}}
\newcommand{\MeV}{\;\mathrm{MeV}}
\newcommand{\TeV}{\;\mathrm{TeV}}
\newcommand{\GeV}{\;\mathrm{GeV}}
\newcommand{\eV}{\;\mathrm{eV}}
\newcommand{\cm}{\;\mathrm{cm}}
\newcommand{\s}{\;\mathrm{s}}
\newcommand{\sr}{\;\mathrm{sr}}
\newcommand{\lab}{\mathrm{lab}}
\newcommand{\ts}{\textstyle}
\newcommand{\ol}{\overline}
\newcommand{\be}{\begin{equation}}
\newcommand{\ee}{\end{equation}}
\newcommand{\ba}{\begin{eqnarray}}
\newcommand{\ea}{\end{eqnarray}}
\newcommand{\nn}{\nonumber}
\newcommand{\nm}{{\nu_\mu}}
\newcommand{\pp}{$\overline{p}(p)-p\;\;$}
\renewcommand{\floatpagefraction}{1.}
\renewcommand{\topfraction}{1.}
\renewcommand{\bottomfraction}{1.}
\renewcommand{\textfraction}{0.}               
\renewcommand{\thefootnote}{F\arabic{footnote}}
\title{Direct CP Violation in $B^\mp \to \pi^\mp \omega, \pi^\mp\rho^0, \pi^0\rho^\mp$, and
in $\ol{B^0}(B^0)\to \pi^\mp\rho^\pm$ With an Enhanced Branching Ratio for $\pi^0\rho^0$}
\author{Saul Barshay$^1$, Georg Kreyerhoff$^1$, and Lalit~M.~Sehgal$^2$\\
\\$^1$III. Physikalisches Institut, RWTH Aachen\\
$^2$ Institut f\"ur theoretische Physik, RWTH Aachen\\
D-52056 Aachen\\Germany}
\maketitle
\begin{abstract}                
We present a novel dynamics for generating sizable CP-violating asymmetries
in the decays of charged $B^\mp\to \pi^\mp \omega, \pi^\mp\rho^0, \pi^0\rho^\mp$, and
in $\ol{B^0}(B^0)\to\pi^\mp\rho^\pm$.
The dynamics for the necessary final-state interactions involves the mixing
of G-parity eigenstates of the system $(\ol{D}^*D,D^*\ol{D})$ with the $G=\pm 1$
states of $\pi\omega$ and $\pi\rho$, respectively. The dynamical effect is enhanced
by the empirically large branching ratio for decays to $(\ol{D}^*D,D^*\ol{D})$.
A correlated result is a markedly enhanced branching ratio for $\ol{B^0}(B^0)\to\pi^0\rho^0$,
which has now been observed in two experiments.
\end{abstract}

Direct CP violation in the decays of charged and neutral $B$ mesons is the central
theme in current experiments \cite{ref1,ref2,ref3} at the two B-meson factories.
Today, some forty years after the discovery of indirect CP violation  in the
two-pion decay of $K^0_L$ \cite{ref4}, direct CP violation has been established 
only in the matrix elements for the two-pion decays of the neutral K system.\cite{ref5}
It is yet to be established in decays of a charged particle. Recently, one
experiment \cite{ref1} has given results which indicate a sizable CP-violating asymmetry
in the decays $B^\mp \to \pi^\mp \eta$, as predicted by theoretical estimates
in 1991 \cite{ref6}, and also in the decays $B^\mp\to K^\mp \eta$.\cite{ref7} Further,
one experiment \cite{ref2} has given a large direct CP violation in $\ol{B^0}(B^0)\to \pi^-\pi^+$.\cite{ref8}
This group has now given an indication \cite{ref3} of a sizable asymmetry in $B^\mp\to\pi^\mp\omega$.
All of these decays have similar, low branching ratios measured to be in the range of
$(2-7)\times 10^{-6}$. In order to have direct CP violation observable, there must
be (strong) interactions among particles in the final states.\cite{ref6} It is physically
clear that if there exists a decay channel with an empirically large branching ratio
(i.~e.~a large decay amplitude) which has the same, conserved strong-interaction quantum
numbers as the final hadron state, then decay into this channel followed by even a small
mixing with the final state, will produce the essential strong-interaction, imaginary
contribution to the amplitude, 
which will be sizable.\cite{ref8} When the large decay amplitude involves a term in 
the CKM matrix with a different weak phase from the term relevant to the
direct decay into the final state, then the necessary conditions for observing an asymmetry
are met.\cite{ref6} This dynamical mechanism explains \cite{ref8} the large, direct CP violation 
observed in $\ol{B^0}(B^0)\to \pi^+\pi^-$.\cite{ref2} In a correlated way, the same dynamics
predicts an enhanced branching ratio for $\ol{B^0}(B^0)\to\pi^0\pi^0$, as is observed \cite{ref9,ref10}.
Mixing with the isospin-zero state of the $\pi\pi$ system occurs from the isospin-zero state of
$D\ol{D}$. The $D^+D^-$ decay mode has a branching ratio now known to be large, 
$\sim 2.5\times 10^{-4}$.\cite{ref11}
In addition to the above-mentioned possible asymmetry in $\pi^\mp\omega$ \cite{ref3},
recent results from the other experiment \cite{ref12}, allow for significant asymmetries with
a definite pattern of signs in 
$B^\mp\to \pi^\mp \rho^0$ and $\pi^0\rho^\mp$. In this paper, we show how sizable asymmetries can
occur in $\pi^\mp\omega$, $\pi^\mp\rho^0$, $\pi^0\rho^\mp$, due to small, strong-interaction mixings
with states of $\ol{D}^*D$ and $D^*\ol{D}$. It is also known that neutral B decay to ${D^*}^\mp D^\pm$
has a large branching ratio, $\sim 8.8\times 10^{-4}$.\cite{ref13} In this analysis, we use an
idea put forward some time ago \cite{ref14}, concerning the presence in charged B 
decays of two distinct,
strong-interaction eigenstates of the charged systems $(\ol{D}^* D$, $D^*\ol{D})$, with different
G-parities, $G=\pm 1$, which are the G-parities of the charged $\pi\omega$ and $\pi\rho$ systems,
respectively. A further striking consequence of the final-state mixing is a marked enhancement of the
branching ratio for $\ol{B^0}(B^0)\to \pi^0\rho^0$, like that for $\pi^0\pi^0$.\cite{ref8}

The following states have $G=\pm 1$, respectively \cite{ref14}. Both charged states have
isospin $I=1$, and the same spin-parity ($0^-$).
\ba
(\ol{D}^* D)_+ &=& \frac{1}{\sqrt{2}}\left({D^*}^- D^0 + {D^*}^0 D^-\right)\;\;\; G=+1, I=1\nn\\
(\ol{D}^* D)_- &=& \frac{1}{\sqrt{2}}\left({D^*}^- D^0 - {D^*}^0 D^-\right)\;\;\; G=-1, I=1
\ea
When mixing of the $G=+1$ state with the state of $\pi^-\omega$ occurs, the physical decay
amplitudes $A$ including final-state interactions, are given in terms of the ``bare'' decay
amplitudes $\tilde{A}$, by
\be
\left( \begin{array}{c}A_{\pi^-\omega}\\
		  A_{(\ol{D}^* D)_+}
	\end{array}\right) =
\left( \begin{array}{cc} \cos\theta_+ & i\sin\theta_+ \\
		   i\sin\theta_+ & \cos\theta_+ 
	\end{array}\right)
\left( \begin{array}{c}\tilde{A}_{\pi^-\omega}\\
                   \tilde{A}_{(\ol{D}^* D)_+}
	\end{array}\right)
\ee
The matrix parameterized by the mixing angle $\theta_+$, is simply the square root of the
S-matrix with neglect of phase factors associated with elastic scattering \cite{ref8}.
The latter were found to have little effect upon calculated asymmetries in $\ol{B^0}(B^0)\to\pi\pi$, 
(as is illustrated in Table 1
and Fig.~1 of Ref.~8). The states of the charged $\pi\rho$ system
with isospin $I=1,2$ are
\be
\left.
\begin{array}{lcrc}
(\pi\rho)_1 &=& \frac{1}{\sqrt{2}}\left(\pi^0\rho^- - \pi^-\rho^0\right) & I=1\\
(\pi\rho)_2 &=& \frac{1}{\sqrt{2}}\left(\pi^0\rho^- + \pi^-\rho^0\right) & I=2
\end{array}
\right\} \;\;G=-1
\ee
The $I=1$ state mixes with the $G=-1$ state $(\ol{D}^*D)_-$ given in Eq.~(1), leading
to physical decay amplitudes given in terms of an angle $\theta_-$ and bare decay
amplitudes by
\be
\left( \begin{array}{c}A_{(\pi\rho)_1}\\
                  A_{(\ol{D}^* D)_-}
        \end{array}\right) =
\left( \begin{array}{cc} \cos\theta_- & i\sin\theta_- \\
                   i\sin\theta_- & \cos\theta_-
        \end{array}\right)
\left( \begin{array}{c}\tilde{A}_{(\pi\rho)_1}\\
                   \tilde{A}_{(\ol{D}^* D)_-}
        \end{array}\right)
\ee
The $\pi\rho$ amplitude with $I=2$ has no mixing; it is given by the bare amplitudes
\be
A_{(\pi\rho)_2} = \tilde{A}_{(\pi\rho)_2}
\ee
Solving Eqs.~(2-5), we obtain three complex, physical decay amplitudes in terms of
two parameters $\theta_+$, $\theta_-$. Since our numerical results are for $|\theta_\pm|\ll 1$,
we simplify the formulae with $\cos\theta_\pm \sim 1$.
\ba
A_{\pi^-\omega} = \tilde{A}_{\pi^-\omega} + i\frac{\sin\theta_+}{\sqrt{2}}\left(
\tilde{A}_{{D^*}^-D^0} + \tilde{A}_{{D^*}^0D^-}\right)\nn\\
A_{\pi^0\rho^-} = \tilde{A}_{\pi^0\rho^-} + i\frac{\sin\theta_-}{2}\left(
\tilde{A}_{{D^*}^-D^0} - \tilde{A}_{{D^*}^0D^-}\right)\\
A_{\pi^-\rho^0} = \tilde{A}_{\pi^-\rho^0} - i\frac{\sin\theta_-}{2}\left(
\tilde{A}_{{D^*}^-D^0} - \tilde{A}_{{D^*}^0D^-}\right)\nn
\ea
In order to show the correlated results for different asymmetries, with no parameters other
than the final-state mixings, 
we use, as we have done in our work on $\ol{B^0}(B^0)\to\pi\pi$
\cite{ref8}, bare amplitudes given by the Bauer-Stech-Wirbel phenomenological model.\cite{ref15}
This model is useful as a first approximation for parameterizing and correlating branching ratios.
\cite{ref16,ref17}
\ba
\tilde{A}_{\pi^-\omega}&\cong& \frac{N\lambda_u}{\sqrt{2}}\left(r'a_1+r a_2\right)\nn\\
\tilde{A}_{\pi^-\rho^0}&\cong& \frac{N\lambda_u}{\sqrt{2}}\left(r'a_1+r a_2\right)\\
\tilde{A}_{\pi^0\rho^-}&\cong& \frac{N\lambda_u}{\sqrt{2}}\left(ra_1+r'a_2\right)\nn
\ea
with parameters $a_1\cong 1$, $a_2\cong 0.2$, $N\cong 0.75$, and the CKM factor $\lambda_u=3.6
e^{-i\gamma}\times 10^{-3}$, with $\gamma\sim 60^\circ$. 
The overall normalization factor $N$ was determined from the
branching ratio for $B^\mp\to\pi^\mp\pi^0$.\cite{ref8} The coefficients $r$ and $r'$ arise from
from ratios of decay constants and overlaps. In the notation of Ref.~\cite{ref17}, they are given by
$r=A_{\pi\rho}/A_{\pi\pi} \sim A_{\pi\omega}/A_{\pi\pi}$ and $r'=A_{\rho\pi}/A_{\pi\pi}\sim
A_{\omega\pi}/A_{\pi\pi}$, and have the approximate values $r\sim 3/2$, $r'\sim 5/4$.
These phenomenological
parameters have at least 15\% uncertainty.
We shall see in Table 1, that these amplitudes produce an adequate first approximation to the
empirically similar charged-particle branching ratios (as do also the related bare amplitudes
for $\ol{B^0}(B^0)\to \pi^-\rho^+, \pi^+\rho^-$ decays discussed below). (The absolute square
of an amplitude gives the branching ratio.) To obtain the bare amplitudes for the coherent states
$(\ol{D}^* D)_\pm$, we use a branching ratio approximately the same as the empirical branching
ratio \cite{ref13} for neutral $B$ decay to ${D^*}^\mp D^\pm$ of $\sim 8.8\times 10^{-4}$, and 
the ratio $|\tilde{A}_{{D^*}^0D^-}/
\tilde{A}_{{D^*}^-D^0}|\sim 0.8$, estimated in Table 7 of Ref.~\cite{ref16}.\cite{ref18} 
This implies a ratio $|(\tilde{A}_{(\ol{D}^*D)_-}/\tilde{A}_{(\ol{D}^*D)_+}|\sim 0.11$ for the
two $G$-parity eigenstates, and allows us to write
\ba
\tilde{A}_{{D^*}^- D^0} \cong 2.62 a_1 \lambda_c,&& \tilde{A}_{{D^*}^0 D^-} \cong 2.10 a_1 \lambda_c\nn\\
\tilde{A}_{{(\ol{D}^*D)}_+} &\cong& 4.72 \frac{a_1\lambda_c}{\sqrt{2}}\\
\tilde{A}_{{(\ol{D}^*D)}_-} &\cong& 0.52 \frac{a_1\lambda_c}{\sqrt{2}}\nn 
\ea
The CKM factor $\lambda_c\cong -8.8\times 10^{-3}$.

Before discussing the results in Table 1, we obtain the amplitude for an additional decay
$\ol{B^0}(B^0)\to\pi^0\omega$, which follows from the same physics as we have described above.
The state which mixes into $\pi^0\omega$, with $\theta_+$, is
\be
\frac{1}{2}\left\{ ({D^*}^-D^+ + {{D^*}^+}D^-) - (\ol{{D^*}^0} D^0 + {D^*}^0\ol{D^0})\right\}
\ee
with $I=1$ and charge-conjugation $C=-1$, like $\pi^0\omega$. The physical decay amplitude
is then
\be
A_{\pi^0\omega}\cong \tilde{A}_{\pi^0\omega} + i\frac{\sin\theta_+}{2}\left( \tilde{A}_{{D^*}^-D^+}
+\tilde{A}_{{D^*}^+D^-}\right)
\ee
where we have taken $\tilde{A}_{\ol{{D^*}^0}D^0}\cong 0 \cong \tilde{A}_{{{D^*}^0}\ol{D^0}}$.\cite{ref15}
We use the approximate bare amplitude from the phenomenological model \cite{ref15,ref16,ref17}
\ba
\tilde{A}_{\pi^0\omega} &\cong& \frac{N\lambda_u}{2}\left(r-r'\right)a_2\nn\\
\mbox{and}\;\; \tilde{A}_{({D^*}^-D^+)_{C=-1}} & \cong& -4.72 \frac{|\lambda_c|}{2}
\ea
from Eqs.~(8,9). From Eqs. (10,11), one immediately observes the same general physical effect as occurred in our
study \cite{ref8} of $\ol{B^0} (B^0)\to \pi^0\pi^0$: although the all-neutral branching ratio is depressed
in the bare amplitude by the small parameter $a_2$, and here also 
by destructive interference, an enhancement can
occur due to the final-state interaction, where a system with a large decay amplitude $|\tilde{A}_{({D^*}^-D^+)_{C=-1}}|$
mixes into the $\pi^0\omega$ final state. 

We calculate direct CP-violating asymmetries from $A_{CP}=(|R|^2-1)/(|R|^2+1)$, where $R$
is the ratio of an amplitude in Eqs.~(6,10) to the amplitude with $\lambda_u\to \lambda_u^*$,
$\lambda_c\to \lambda_c^*$.
In Table 1, we give representative asymmetries and branching ratios for the three charged decay
modes and for $\pi^0\omega$, calculated from the amplitudes in Eqs.~(6,10) using small mixing
angles $\theta_+=+0.025$ and $\theta_-=0.25$. (A single mixing of order 0.2
occurred in our work on
$\ol{B^0}(B^0)\to \pi\pi$.\cite{ref8}) We tabulate recent experimental results, from which it
is clear that there are indications of possible sizable asymmetries with a definite pattern
of signs. However, in
contrast to this possibility which is given by our results using final-state interactions
among specific systems of physical hadrons, there are representative results from recent calculations
which neglect final-state interactions of this kind. In Table 1 we also list these asymmetry results
\cite{ref17,ref19} (for wide variations in these estimates, see the tabulations in Ref.~17).
Generally, the asymmetries in these calculations are small because of small strong-interaction phases. 
 
There are striking distinctions in the $\pi\omega$ system, in particular. The mixing to
$(\ol{D}^*D)_+$, even with the small mixing angle $\theta_+=0.025$, can result in a sizable asymmetry $A_{CP}(\pi^\mp\omega)\sim +0.4$, in
contrast to the very small $\sim -0.02$ in the last column of Table 1. The calculated branching
ratio for $\pi^0\omega$ is about 5\% of the branching ratio for $\pi^-\omega$, instead of the
miniscule $\sim 0.1\%$ given in Table 9 of Ref.~17. The elevated branching ratio is directly
correlated with the possible sizable asymmetry in $\pi^\mp\omega$. Both effects arise from
mixing with the $(\ol{D}^*D,D^*\ol{D})$ system.
Since the $\pi^0\omega$ amplitude is dominated  by the term from the strong-interaction
mixing, 
the parameter for indirect CP violation \cite{ref8}, $S_{\pi^0\omega}\sim -\sin 2\beta \cong -0.7$ 
 (for $2\beta\cong 45^\circ$).

To complete the results which follow from this dynamics for final-state interactions, we give
estimates for asymmetries and branching ratios in three more related decays $\ol{B^0}(B^0)\to
\pi^-\rho^+$, $\pi^+\rho^-$, $\pi^0\rho^0$. Significant asymmetries may be present in the
$\pi^\mp\rho^\pm$ modes.\cite{ref20,ref21,ref22} In addition to the $I=1,2$ states, there is now an
$I=0$ state. The isospin states with $I_3=0$ are
\ba
(\pi\rho)_0 &=& \frac{1}{\sqrt{3}}\left(\pi^-\rho^+ + \pi^+\rho^- - \pi^0\rho^0\right)\nn\\
(\pi\rho)_1 &=& \frac{1}{\sqrt{2}}\left(\pi^+\rho^- - \pi^-\rho^+\right)\\
(\pi\rho)_2 &=& \frac{1}{\sqrt{6}}\left(\pi^-\rho^+ + \pi^+\rho^- + 2\rho^0\pi^0\right)\nn
\ea
The $I=0$ state with $C=-1$, can mix via a third angle $\theta$, with the state
\be
\frac{1}{2}\left\{ \left( {D^*}^- D^+ + {D^*}^+ D^-\right) + \left(\ol{{D^*}^0} D^0 + {D^*}^0 \ol{D^0}\right)\right\}
\ee
Together with Eq.~(4,5), one solves for the three additional physical decay amplitudes.
\ba
A_{\pi^-\rho^+}&=& \tilde{A}_{\pi^-\rho^+} - i\frac{\sin\theta_-}{2\sqrt{2}}\left( \tilde{A}_{{D^*}^-D^+} -
\tilde{A}_{{D^*}^+ D^-}\right) \nn\\
&& +i\frac{\sin\theta}{2\sqrt{3}}\left(\tilde{A}_{{D^*}^-D^+} +
\tilde{A}_{{D^*}^+D^-}\right)\nn\\
A_{\pi^+\rho^-}&=& \tilde{A}_{\pi^+\rho^-} + i\frac{\sin\theta_-}{2\sqrt{2}}\left( \tilde{A}_{{D^*}^-D^+} -
\tilde{A}_{{D^*}^i+ D^-}\right)\\
&& + i\frac{\sin\theta}{2\sqrt{3}}\left(\tilde{A}_{{D^*}^-D^+} +
\tilde{A}_{{D^*}^+D^-}\right)\nn\\
A_{\pi^0\rho^0}&=& \tilde{A}_{\pi^0\rho^0} - i\frac{\sin\theta}{2\sqrt{3}}\left( \tilde{A}_{{D^*}^-D^+} +
\tilde{A}_{{D^*}^+ D^-}\right)\nn 
\ea
As in Eqs.~(7,11), we use phenomenological, model amplitudes \cite{ref16,ref17} as a first approximation
\ba
\tilde{A}_{\pi^-\rho^+}&\cong& N\lambda_u r' a_1\nn\\
\tilde{A}_{\pi^+\rho^-}&\cong& N\lambda_u r a_1\\
\tilde{A}_{\pi^0\rho^0}&\cong& \frac{N\lambda_u}{2}\left(r+r'\right)a_2\nn
\ea
Note that the five complex, physical amplitudes in the $\pi\rho$ system satisfy a ``pentagon'' relationship \cite{ref23}
(as do the bare amplitudes, of course)
\be
\frac{A_{\pi^-\rho^+}+A_{\pi^+\rho^-}}{2}+A_{\pi^0\rho^0} = \frac{A_{\pi^-\rho^0}+A_{\pi^0\rho^-}}{\sqrt{2}}
\ee
The additional asymmetries and branching ratios are given in Table 1, as calculated with $\theta=-0.09$. There
is a suggestion in the recent data \cite{ref20,ref21,ref22} on $\pi^-\rho^+$, $\pi^+\rho^-$ and
$\pi^\mp\rho^\pm$ of  
significant asymmetries. Our results give this, in $\ol{B^0}\to\pi^-\rho^+$ in particular.\cite{ref21}
Our estimated asymmetry without flavor tagging ($\pi^\mp\rho^\pm$) agrees with the data.\cite{ref22}
For comparison, the asymmetries
indicated in the last column of Table 1 \cite{ref17} are very small, as they must be when there
is little strong interaction. The separate branching ratios for $\pi^-\rho^+$ and $\pi^+\rho^-$
come closer together as a consequence of the final-state interactions, in agreement with indications
from the data.\cite{ref22} Our estimated branching ratio for $\pi^0\rho^0$ is enhanced.
Enhancement is observed in the data.\cite{ref12,ref24}
The size is essentially determined by the mixing $\theta$ with the coherent $C=-1$ $(\ol{D}^*D,D^*\ol{D})$
system which is produced with a large amplitude. Physically, this is the same dynamics that gives
rise to the enhanced branching ratio of $\sim 2\times 10^{-6}$ for $\ol{B^0}(B^0)\to \pi^0\pi^0$.\cite{ref8}
The calculated parameter $S_{\pi^0\rho^0}\cong -0.56$;  it differs from $-\sin(2\beta+2\gamma)\cong
-0.26$ because of the final-state interactions.

It is to be expected that $\ol{B^0}(B^0)\to \eta\omega$, will obtain a contribution from mixing
with the same $(\ol{D}^*D,D^*\ol{D})$ system as $\pi^0\rho^0$, and thus will have a 
similar, enhanced branching 
ratio. This enhancement has just been observed in one experiment.\cite{ref25} 

Our present results on possible sizable asymmetries in certain charged-B decays are
calculated using a physical-hadron approach to estimating final-state interactions, which has been
successful.\cite{ref8,ref6} There are present experimental indications that specific,
charged-decay modes, $B^\mp\to \pi^\mp\omega$, $\pi^\mp\rho^0$, $\pi^0\rho^\mp$, \cite{ref3,ref12},
and $\ol{B^0}(B^0)\to
\pi^-\rho^+$, $\pi^+\rho^-$, $\pi^\mp\rho^\pm$,\cite{ref20,ref21,ref22},
have significant asymmetries. We have given results for all of the
asymmetries. For chosen small mixing angles, these results exhibit distinctive correlations
in sign and in magnitude, which follow the pattern of the present data. Our positive results
for these modes should encourage the experimenters to pursue these asymmetries, in the effort
to finally establish CP violation in charged-particle decays. Directly correlated with the
final-state interactions which give rise to asymmetries is the
dynamical enhancement of decays: $\ol{B^0}(B^0)\to \pi^0\pi^0,\;\pi^0\rho^0,\;\pi^0\omega$. Striking
enhancement of the $\pi^0\pi^0$ and $\pi^0\rho^0$ rates has appeared in the recent
data.\cite{ref9,ref10,ref12,ref24}

\newpage
\begin{landscape}
\begin{table}
\begin{tabular}{l|r|r|r|r|r|r|r|r|c|r}
&\multicolumn{3}{c|}{ $Br(\ol{B}/B)$} & \multicolumn{3}{c|}{$A_{CP}$}  & \multicolumn{3}{c|}{data}    & Ref.~17\\
\hline
&  $\theta_i=0$ & $\theta_i\neq 0$  & $N_\omega=0.6$ & $\theta_i=0$  & $\theta_i\neq 0$      &  $N_\omega=0.6$ & $Br(\ol{B}/B)$ &  $A_{CP}$ & Ref. & $A_{CP}$ \\
\hline
\hline
$\pi^-\omega$ & 8.8 & 9.3 & 6.1 & 0 & +0.40 & +0.49 & $5.7^{+1.4}_{-1.3}\pm 0.6$ & $+0.50^{+0.25}_{-0.21}\pm 0.02$ & \cite{ref3} & $-0.02$ \\
              &     &     &     &   &       &       & $5.5\pm 0.9 \pm 0.5$ & $+0.03\pm 0.16\pm 0.03$ & \cite{ref1} & \\
\hline
$\pi^-\rho^0$ & 8.8 & 9.1 &  & 0 & $-0.32$ &  & $9.5\pm 1.1 \pm 0.8$ & $-0.19\pm 0.11\pm 0.02$ & \cite{ref12} & +0.04 \\
\hline
$\pi^0\rho^-$ & 11.2 & 11.5 &  & 0 & $+0.28$ &  & $10.9\pm 1.9 \pm 1.9$ & $+0.24\pm 0.16\pm 0.06$ & \cite{ref12} & $-0.04$ \\
 &  &  &  &  &  &  & $13.8\pm 2.4^{+1.5}_{-1.6}$ & $+0.06\pm 0.19\pm 0.04$ & \cite{ref24} &  \\
\hline
$\pi^0\omega$ & $0.45\times 10^{-2}$& 0.27 & 0.27 & 0 & +0.22 & +0.18 & $<1.9$ &  & \cite{ref3} & \\
\hline
$\pi^\mp\rho^\pm$ & 27.8 &30.4 & & 0 & $-0.13$ & & $22.6\pm 1.8\pm 2.2$ & $-0.18\pm 0.08 \pm 0.03$ & \cite{ref20,ref21} & \\
                  &      &      & & &        & &                   & $-0.114\pm 0.062 \pm 0.027$ & \cite{ref22} & \\
\hline
$\pi^-\rho^+$ & 11.4 & 13.6 &    & 0 & $-0.63$ & &                   &                        &              & +0.006\\
\hline
$\pi^+\rho^-$ & 16.4 & 16.8 &      & 0 & $-0.28$ & &                    &                      &              & $-0.015$\\
\hline
$\pi^0\rho^0$ & 0.55 & 1.71 &        & 0 & $+0.81$ & &  $1.4\pm 0.6\pm 0.3$ &                       & \cite{ref12} & $-0.16$\\
	  &        &     &        &   &       & & $5.1\pm 1.6\pm 0.8 $&                            & \cite{ref24} &           
\end{tabular}
\caption{The flavor-averaged branching ratio $Br(\ol{B}/B)$ in units of $10^{-6}$, and the
asymmetry $A_{CP}$\cite{ref3,ref20}, as calculated from the amplitudes in Eqs.~(6,10,14). The first
columns under $Br$ and $A_{CP}$ have $\theta_+=\theta_-=\theta=0$; the second columns $\theta_+=0.025$,
$\theta_-=0.25$, $\theta=-0.09$. The numbers under $N_\omega=0.6$ are for $N\to N_\omega$ in the
$\pi\omega$ bare amplitudes in Eqs.~(7,11). We have not listed data for $\pi^-\rho^+$ and $\pi^+\rho^-$
separately. These numbers involve correlated data parameters. The separation is not given explicitly
in Refs.~(20,22). See p.~10 of Ref.~21 for separate asymmetries from the data. These do follow the
distinctive pattern of those calculated in this paper and listed above, with $A_{CP}(\pi^-\rho^+)\sim
-0.62\pm 0.27 < A_{CP}(\pi^+\rho^-)\sim -0.11\pm 0.17$.}
\end{table}
\end{landscape}
\end{document}